\pdfoutput=1
\documentclass[letter]{aa}
\usepackage{linenoaa}
\usepackage{graphicx}
\usepackage{txfonts}
\usepackage{natbib}
\usepackage{xcolor}
\usepackage{booktabs}
\usepackage{url}
\bibpunct{(}{)}{;}{a}{}{,}
\usepackage{hyperref}
\hypersetup{pdfencoding=auto,colorlinks=true,citecolor=blue!60!black,linkcolor=blue!60!black,urlcolor=blue!60!black,
  pdftitle={}, pdfauthor={}, pdfsubject={}, pdfkeywords={}, pdfcreator={}, pdfproducer={}}
\pdftrailerid{}
\newcommand{\kms}{\mathrm{km\,s^{-1}}}
\newcommand{\hmpc}{h^{-1}\,\mathrm{Mpc}}

\begin{document}

\title{Physical quasar pairs in DESI DR1 as tracers of protoclusters at $z\approx2$}
\author{Mart\'{\i}n G. Richarte\inst{1,2}\thanks{\email{martin@df.uba.ar} (Corresponding author)} \and Diego G. Lambas\inst{3,4}\thanks{\email{diego.garcia.lambas@unc.edu.ar}}}
\institute{Departamento de F\'{\i}sica, Facultad de Ciencias Exactas y Naturales, Universidad de Buenos Aires, Buenos Aires, Argentina
\and Centro de Ci\^encias Naturais e Humanas, Universidade Federal do ABC, Santo Andr\'e, SP, Brazil
\and Instituto de Astronom\'{\i}a Te\'orica y Experimental (IATE), CONICET-UNC, C\'ordoba, Argentina
\and Observatorio Astron\'omico de C\'ordoba (OAC), UNC, C\'ordoba, Argentina}
\date{}

\abstract
{Close pairs of quasars at the same redshift are expected to mark overdense regions that will collapse into galaxy clusters. Earlier work with SDSS found about 2800 such pairs, too few to follow their environment with separation.}
{We build the largest sample of physical quasar pairs to date from DESI DR1 and test whether pairs out to Mpc separations trace protoclusters.}
{We select pairs with rest-frame velocity difference $|\Delta V|\leq900~\kms$ and projected separation $r_p\leq4$~Mpc at $1.2\leq z\leq2.8$, measure their purity and fibre completeness, and measure the clustering of pair members around all DESI quasars.}
{We find 40\,626 physical pairs with projected separation $r_p\leq2.5$~Mpc, where their purity falls to 50\%, the effective radius of protoclusters at $z\approx2$. Pair members are $1.69\pm0.14$, $1.78\pm0.09$ and $2.01\pm0.06$ times more clustered than all quasars for projected separations $r_p\leq1$, 1--2.5 and 2.5--4~Mpc, while chance pairs show little excess ($1.08\pm0.05$). This corresponds to haloes of $\sim1.4$--$2.2\times10^{13}\,h^{-1}M_\odot$.}
{Physical quasar pairs with projected separations up to 2.5~Mpc trace protocluster environments, as predicted by the AbacusSummit $\Lambda$CDM mocks. The full DESI survey will turn them into a statistical census of protoclusters at cosmic noon.}


\maketitle
\nolinenumbers

\begin{figure*}
\centering
\includegraphics[width=0.74\textwidth]{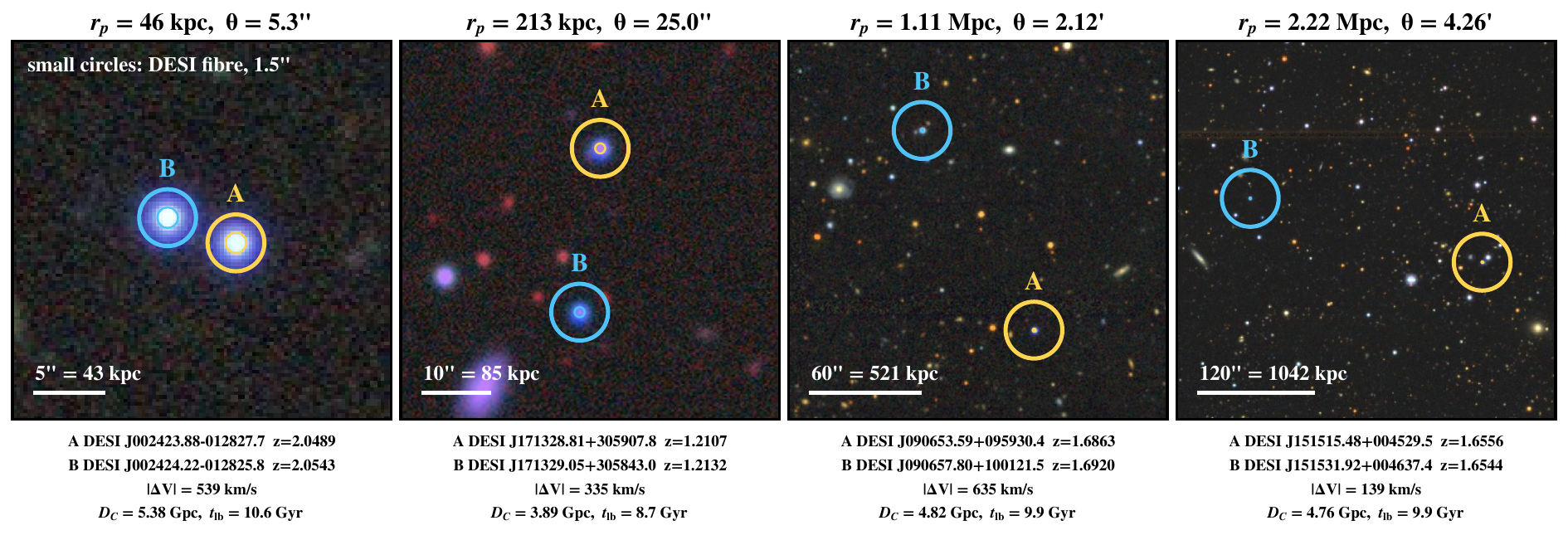}
\caption{Four physical quasar pairs of the DESI DR1 sample (46~kpc to 2.2~Mpc) on Legacy Survey DR10 $grz$ images. Yellow and blue circles: members A and B; small circles: 1.5\arcsec\ DESI fibre. Labels give names, redshifts, $|\Delta V|$, comoving distance $D_C$ and lookback time $t_{\rm lb}$; scale bars in physical kpc. North up, east left. Spectra in Fig.~\ref{fig:spectra}.}
\label{fig:examples}
\end{figure*}

\section{Introduction}

The progenitors of galaxy clusters at $z\gtrsim2$, the protoclusters, are where much of the stellar mass of today's massive galaxies formed \citep{Overzier2016}. At $z\approx2$ they are not yet virialised and extend over effective radii of 5--10 comoving Mpc, with members spread over several haloes \citep{Chiang2013,Muldrew2015}, so most were found around rare tracers in small samples \citep{Overzier2016}.

Quasars are visible to $z>2$ over large areas, have spectroscopic redshifts, and live in haloes of a few $10^{12}\,h^{-1}M_\odot$ \citep{Laurent2017}. A single quasar marks only a modest overdensity, but two quasars close on the sky and in redshift are expected where the density of massive haloes is highest. Their small-scale excess has long been known \citep{Hennawi2006}, individual systems such as the quasar quartet of \citet{Hennawi2015} sit in protoclusters, and the strongest quasar overdensities at $z\approx2$ now test $\Lambda$CDM \citep{Kuwayama2026}.

Close quasar pairs could thus trace dense environments statistically, given a large spectroscopic sample. Using the DR16 quasar catalogue, \citet[][hereafter MR25]{Richarte2025} identified 2777 physical quasar pairs with $|\Delta V|\leq900~\kms$ and projected separations $d_p\leq2$~Mpc at $1.2\leq z\leq2.8$. With pair--quasar cross-correlations and Planck CMB lensing, they showed that pairs are more strongly biased than single quasars. How this signal evolves with separation, and where the physical association ends, requires more pairs than SDSS can provide.

The Dark Energy Spectroscopic Instrument (DESI) now makes this step possible. Its first data release (DR1) contains more than a million quasars at $z>1$ \citep{DESIDR1}, with dense coverage from repeated passes over the same sky. \citet{Jing2026} compiled 1842 pair candidates closer than 110~kpc, and \citet{He2025} identified lensed, dual and projected quasars in the DESI Legacy Imaging Surveys. \citet{Hada2026} used $\sim10^4$ foreground/background pairs to measure transverse proximity zones. \textit{Gaia} astrometry offers another route to close pairs \citep{Chen2026Gaia}. Physical pairs at Mpc separations, and their environment, remain to be explored.

In this Letter we use DESI DR1 to test whether physical quasar pairs trace protoclusters, and we revisit each step of MR25: we adopt the rest-frame velocity difference, set the outer separation limit from the measured purity, correct for fibre assignment, add a chance-pair control sample, and measure the clustering of pair members as a function of their separation (Fig.~\ref{fig:examples}). We adopt the Planck 2018 cosmology \citep{Planck2018}. Separations $r_p$ are in physical Mpc unless written in $\hmpc$, which are comoving.

\section{Data}
\label{sec:data}

\subsection{DESI DR1 quasars}
DESI is a 5000-fibre spectrograph on the 4-m Mayall telescope at Kitt Peak \citep{DESI2016,DESI2022}. Quasar targets are selected from the Legacy Imaging Surveys \citep{Chaussidon2023}, and redshifts come from the pipeline \citep{Guy2023} with \textsc{Redrock}, \textsc{QuasarNET} and a Mg\,\textsc{ii} classifier \citep{Alexander2023}. DR1 contains the first 13 months of the main survey and is public \citep{DESIDR1}.\footnote{\url{https://data.desi.lbl.gov/doc/releases/dr1/}}

We start from the DR1 quasar catalogue (\texttt{QSO\_cat\_iron\_cumulative\_v0}; 2\,003\,848 spectra).\footnote{\url{https://data.desi.lbl.gov/public/dr1/survey/catalogs/dr1/QSO/iron/}} We keep one spectrum per \texttt{TARGETID} (main dark survey first, then highest \texttt{TSNR2\_QSO}; 1\,500\,452 quasars), require \texttt{ZWARN}~$=0$ and \texttt{COADD\_FIBERSTATUS}~$=0$ (1\,432\,062), and remove duplicates closer than 1\arcsec\ (1\,431\,879). As in MR25 we study $1.2\leq z\leq2.8$, where the quasar density is highest (1\,088\,242 quasars); the pair search uses the 1\,183\,335 quasars at $1.1\leq z\leq2.9$, so that pairs with mean redshift inside the range are not lost at the edges.
For the completeness correction and the clustering we also use the DR1 large-scale-structure (LSS) quasar catalogues \citep{Ross2025}. The \texttt{full} catalogue lists all 2\,248\,020 quasar targets, observed or not, with the number of overlapping tiles $N_{\rm tile}$ and the fibre-assignment flags. The clustering catalogues contain 1\,015\,192 quasars at $1.1<z<2.9$ with completeness and systematic weights (\texttt{WEIGHT}), and their randoms, of which we use 6.0 million (40\% of random file 0).\footnote{Version v1.5, \url{https://data.desi.lbl.gov/public/dr1/survey/catalogs/dr1/LSS/iron/LSScats/v1.5/}}

\subsection{Pair selection}
\label{sec:pairs}
With the parent sample defined, we search for pairs among all quasars with a $k$-d tree on the unit sphere. For two quasars with redshifts $z_1$, $z_2$ and mean $\bar z$ we use the rest-frame velocity difference
\begin{equation}
|\Delta V| = \frac{c\,|z_1-z_2|}{1+\bar z},
\end{equation}
and the projected physical separation $r_p=\theta\,D_A(\bar z)$. MR25 used $c\,|z_1-z_2|$ without the $(1+\bar z)$ factor. At $z\approx2$ that limit is three times stricter (12\,838 pairs instead of 28\,819).
We require $|\Delta V|\leq900~\kms$, which contains 77\% of the excess pairs over chance at every $r_p$ (93\% below $1500~\kms$). The width is set by redshift errors rather than peculiar velocities, and corresponds to $\approx13$ comoving Mpc along the line of sight at $z=2$, comparable to the extent of a protocluster. With $r_p\leq4$~Mpc and $1.2\leq\bar z\leq2.8$ we find 84\,372 pairs (28\,819 with $r_p\leq2$~Mpc and 40\,789 with $r_p\leq2.5$~Mpc).
At the smallest separations we use the visually inspected catalogue of \citet{Jing2026} to remove 29 lensed-quasar candidates and 121 pairs with a non-quasar member, all below 110~kpc (Appendix~\ref{app:extra}). We also drop 12 known lenses \citep{Lemon2023} and two pairs that repeat a companion through both images of a lens. This leaves 84\,208 pairs with $r_p\leq4$~Mpc and 40\,626 with $r_p\leq2.5$~Mpc, our main sample (Sect.~\ref{sec:purity}); 28\,656 of them have $r_p\leq2$~Mpc, the limit of MR25.

Figure~\ref{fig:examples} shows four pairs of the final sample, from a close pair resolved in the imaging to a pair 2.2~Mpc apart. An automatic inspection of the images and spectra of 90 random pairs confirms 89 as two distinct quasars and finds no lensed system (Appendix~\ref{app:inspection}).

\begin{figure*}[!t]
\centering
\includegraphics[width=0.82\textwidth]{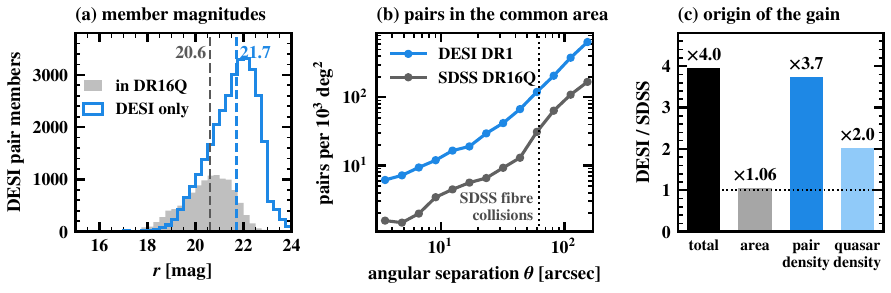}
\caption{DESI DR1 and SDSS DR16Q pairs with $|\Delta V|\leq900~\kms$ and $r_p\leq2$~Mpc. (a) $r$ magnitude of DESI members in the common footprint, also in DR16Q (grey) or only in DESI (blue); dashed lines: medians. (b) Pairs per $10^3$~deg$^2$ in the common footprint versus angular separation; dotted line: 62\arcsec\ SDSS fibre-collision radius. (c) DESI/SDSS ratio: total pairs, area, and pair and quasar densities.}
\label{fig:sdss}
\end{figure*}

Compared with SDSS DR16Q, DESI finds four times more pairs within $r_p\leq2$~Mpc, mainly by reaching fainter quasars rather than a larger volume (Fig.~\ref{fig:sdss}; Appendix~\ref{app:sdss}).

\subsection{Purity and the outer separation limit}
\label{sec:purity}
We count chance pairs by shuffling redshifts among quasars at fixed positions (ten realisations). The purity $P=1-N_{\rm chance}/N_{\rm obs}$ is 0.97 at $r_p<0.06$~Mpc, 0.71 at 0.75--1~Mpc, 0.53 at 2--2.5~Mpc and 0.43 at 3--4~Mpc (Fig.~\ref{fig:purity}). We adopt $r_p=2.5$~Mpc as the outer limit of the main sample, because beyond it more than half of the pairs are chance projections. This limit is set by the data but is consistent with the size of protoclusters: the progenitors of $z=0$ clusters with $M\gtrsim10^{14}M_\odot$ have effective radii $R_e\approx5$--10 comoving Mpc at $z\approx2$, or 1.7--3.3 physical Mpc \citep{Chiang2013}, so members can be $\sim2R_e$ apart. The 2.5--4~Mpc bin is kept as a test.

\subsection{Fibre-assignment completeness}
DESI cannot place fibres on two close targets in a single pass. We measure the pair completeness $f(\theta,N_{\rm tile})$ as the ratio of angular pairs among targets that received a fibre to pairs among all quasar targets of the LSS \texttt{full} catalogue, as a function of the number of overlapping tiles (Appendix~\ref{app:extra}). In single-pass regions $f\approx0.01$--0.04 below 1\arcmin, which no weighting can recover, while for $N_{\rm tile}\geq2$ $f\geq0.67$ at all separations. We therefore use pairs with both members in $N_{\rm tile}\geq2$ regions, and weight pair counts by $1/f$. The purity, a ratio of counts at fixed $\theta$, is unchanged by this weight, and the clustering, a per-member statistic, uses the LSS weights of each member.
Figure~\ref{fig:completeness} shows $f(\theta,N_{\rm tile})$. A test with single quasars confirms that restricting to $N_{\rm tile}\geq2$ regions does not bias the clustering amplitude ($0.98\pm0.02$). Around quasars without a companion, the excess of unobserved quasar targets within 1~Mpc, an upper limit on missed companions, is 0.13 per quasar in single-pass regions but only 0.02 for $N_{\rm tile}\geq2$.

\subsection{Control samples}
We compare pair members with all DESI quasars and with chance pairs with $6000<|\Delta V|\leq12\,000~\kms$ at the same $r_p$, which are $\gtrsim60~\hmpc$ apart along the line of sight, beyond the projection depth of the clustering measurement.

\begin{figure*}[t]
\centering
\includegraphics[width=0.88\textwidth]{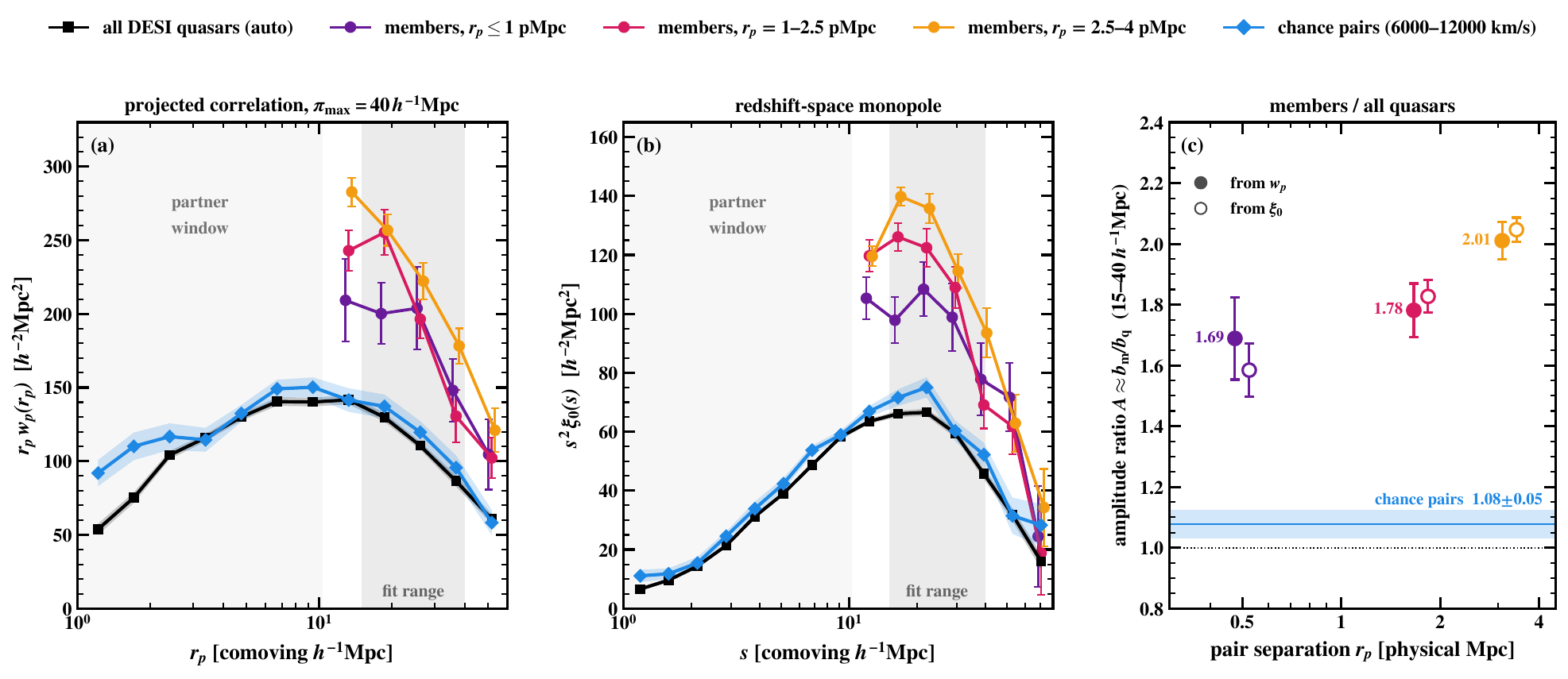}
\caption{Clustering of pair members with all DESI quasars. (a) Projected correlation $r_p w_p$ with $\pi_{\max}=40~\hmpc$. (b) Redshift-space monopole $s^2\xi_0$. Points: members of physical pairs in three $r_p$ bins; lines with $1\sigma$ bands: quasar auto-correlation and chance-pair members. Shaded: 15--40~$\hmpc$ fit range; light region below $10.3~\hmpc$: partner window, where removing the selected partners shapes the member correlations (not shown). (c) Amplitude ratio to all quasars from $w_p$ (filled) and $\xi_0$ (open); blue band: chance-pair value. Errors from 64 jackknife regions.}
\label{fig:clustering}
\end{figure*}

\section{Methods}
\label{sec:methods}

\subsection{Clustering of pair members}
We use each member of a pair as a tracer and cross-correlate it with all LSS quasars, with the Landy--Szalay estimator \citep{Landy1993} in \textsc{pycorr}\footnote{\url{https://github.com/cosmodesi/pycorr}} and the \textsc{Corrfunc} engine \citep{Sinha2020}.\footnote{\url{https://github.com/manodeep/Corrfunc}} For the auto-correlation we use the same data and randoms. We measure $w_p(r_p)$ with $\pi_{\max}=40~\hmpc$ and the redshift-space monopole $\xi_0(s)$. Each quasar enters the tracer sample once, even when it belongs to several pairs of the same bin, and all the partners that satisfy the selection are removed from the data--data counts, in the total and in every jackknife subsample. Otherwise, the companions imposed by the selection would be counted as signal. These companions lie at comoving separations below $10.3~\hmpc$ (4 physical Mpc and $\pi$ at $z=2.8$), so their removal leaves a deficit in the counts at smaller scales, and we fit only above this scale. Errors come from 64 angular jackknife regions. We quote the amplitude ratio
\begin{equation}
A = \frac{\sum_i R_i\,w^{\rm m\times q}(R_i)}{\sum_i R_i\,w^{\rm qq}(R_i)},\qquad 15<R_i<40~\hmpc,
\end{equation}
where $w$ stands for $w_p$ or $\xi_0$. On these scales $A\approx b_{\rm m}/b_{\rm q}$, the ratio of the bias of the members' environment to that of all quasars.

\section{Results}
\label{sec:results}

\subsection{Clustering as a function of pair separation}
Figure~\ref{fig:clustering} and Table~\ref{tab:amp} give the clustering of pair members, chance pairs and all quasars. Members of physical pairs are 1.7--2.0 times more clustered than all quasars, while chance pairs are close to all quasars ($A=1.08\pm0.05$). The amplitude is larger in the outermost bin than in the innermost one, at the $2.1\sigma$ level. The $w_p$ and $\xi_0$ estimates agree within the errors. Splitting the $r_p\leq2.5$~Mpc pairs by velocity difference shows no trend: $A(w_p)=1.71\pm0.11$, $1.77\pm0.12$ and $1.79\pm0.13$ for $|\Delta V|$ of 0--300, 300--600 and 600--900~$\kms$.

For all DESI quasars, a power law $\xi(r)=(r/r_0)^{-\gamma}$ fitted to $w_p$ at 5--25~$\hmpc$ gives $r_0=6.83\pm0.11~\hmpc$ and $\gamma=1.83\pm0.03$ ($\chi^2=1.7$ for 2 degrees of freedom), in line with earlier quasar measurements at $z\approx1.5$--2 \citep{Laurent2017}. Power-law fits to the member correlations, restricted to scales above the partner window, are given in Appendix~\ref{app:extra}.

Table~\ref{tab:pl} gives power-law fits $\xi(r)=(r/r_0)^{-\gamma}$ to $w_p$ at 10--40~$\hmpc$, above the partner window, integrating the model to $\pi_{\max}=40~\hmpc$ and using the jackknife covariance. Members have $r_0=8.6$--$10.8~\hmpc$, against 7.8 for all quasars.

\section{Interpretation}
\label{sec:interp}

To translate the ratios into masses we fit the quasar auto-correlation $w_p$ over 5--40~$\hmpc$ with the linear redshift-space model \citep{Kaiser1987,Hamilton1992}, projected to the same $\pi_{\max}$, at the effective redshift $z_{\rm eff}=1.87$. This gives $b_{\rm q}=2.68\pm0.02$ ($\chi^2=5.0$ for 5 degrees of freedom), close to earlier quasar measurements at this redshift \citep{Laurent2017}. The measured ratios then imply environment biases $b\approx4.5$, 4.8 and 5.4 for the three $r_p$ bins (Table~\ref{tab:amp}). These correspond to haloes of $1.4$, $1.6$ and $2.2\times10^{13}\,h^{-1}M_\odot$ with the \citet{Tinker2010} bias relation, compared with $2.6\times10^{12}\,h^{-1}M_\odot$ for all quasars. At $z\approx1.9$, such regions grow into clusters of $\gtrsim10^{14}M_\odot$ by $z=0$, as expected for protoclusters \citep{Chiang2013}.

To see how much of this excess $\Lambda$CDM predicts, we repeat the measurement on the ten complete DESI DR1 AbacusSummit quasar mocks \citep{Maksimova2021},\footnote{\url{https://abacussummit.readthedocs.io/}; DESI DR1 mocks: \url{https://desidatamodel.readthedocs.io/en/latest/DESI_ROOT/survey/catalogs/RELEASE/mocks/AbacusSummit/index.html}} in which quasars populate haloes following the occupation fitted to the DESI quasar clustering \citep{Yuan2024}. The mocks cover $z\leq2.1$, so we restrict both mocks and data to $1.2\leq z\leq2.1$. Here we use the Davis--Peebles estimator, with the same selection and partner removal in data and mocks. Averaged over the ten realisations, members of pairs with $r_p\leq1$, 1--2.5 and 2.5--4~Mpc are $1.48$, $1.62$ and $1.73$ times more clustered than all quasars, with a scatter of 0.06, 0.04 and 0.03 between realisations, similar to the jackknife errors. The data give $1.70\pm0.16$, $1.60\pm0.09$ and $1.92\pm0.09$ over the same redshift range, that is, data-to-mock ratios of $1.14\pm0.12$, $0.99\pm0.06$ and $1.11\pm0.05$. The largest difference, $2.0\sigma$ in the outermost bin once the mock scatter is included, is not significant. Quasar pairs thus select protocluster-scale environments at the level that $\Lambda$CDM predicts. The growth of the bias with $r_p$ is also seen in every mock. The clustering ratio measures the bias of the environment, not the host-halo mass.

\section{Summary}

We have built the largest sample of physical quasar pairs to date from DESI DR1. Using the rest-frame velocity difference, we select 40\,626 pairs with $|\Delta V|\leq900~\kms$ and projected separation $r_p\leq2.5$~Mpc at $1.2\leq z\leq2.8$. This outer limit is where the purity falls to 50\%, and matches the effective radius of protoclusters at $z\approx2$. Within the 2~Mpc limit of MR25 we find 28\,656 pairs, ten times their SDSS sample; the gain comes from fainter quasars, which double the quasar density, rather than a larger volume.

In regions with at least two DESI passes, pair members are 1.7--2.0 times more clustered than all DESI quasars, while chance pairs show at most a small excess. This corresponds to environments with the bias of $\sim1.4$--$2.2\times10^{13}\,h^{-1}M_\odot$ haloes at $z\approx1.9$, the expected progenitors of present-day clusters, and ten AbacusSummit mocks of the DESI quasars reproduce it within the errors. Physical quasar pairs thus behave as $\Lambda$CDM predicts for tracers of protoclusters.

These results open a clear path forward. The full DESI survey will substantially enlarge the pair sample over the same redshift range and sharpen the trend of the bias with separation. The mass of these environments can also be weighed with CMB lensing: a stacked convergence ($\kappa$) analysis of the pairs with ACT and \textit{Planck} will be presented in a future work. If pairs mark protoclusters, the most massive should also strongly lens background galaxies; to our knowledge this has not been searched for around quasar pairs, and deep \textit{HST}, \textit{JWST} or \textit{Euclid} imaging of the richest pair environments would test it directly. Conversely, a foreground cluster can magnify a protocluster core, as in PJ0846+15 at $z\approx2.66$ \citep{Foo2025}; pairs behind massive clusters would offer the same view. Physical quasar pairs from DESI can therefore become a statistical census of protoclusters at cosmic noon.

\section*{Data availability}
The pair catalogue, completeness weights, measurements and scripts used in this Letter will be made available upon publication, on reasonable request to the first author (M.G.R.).

\begin{acknowledgements}
M.G.R. is partially supported by UFABC (Brazil) and WFSA. This work was partially supported by the Agencia Nacional de Promoci\'on Cient\'{\i}fica y Tecnol\'ogica (PICT 2015-3098, PICT 2016-1975), the Consejo Nacional de Investigaciones Cient\'{\i}ficas y T\'ecnicas (CONICET, Argentina) and the Secretar\'{\i}a de Ciencia y Tecnolog\'{\i}a de la Universidad Nacional de C\'ordoba (SeCyT-UNC, Argentina).

This work uses public data from DESI DR1 and the AbacusSummit simulations.
\end{acknowledgements}

\clearpage
\begin{appendix}
\nolinenumbers
\section{Sample checks and power-law fits}
\label{app:extra}
\begin{figure}[!ht]
\centering
\includegraphics[width=\columnwidth]{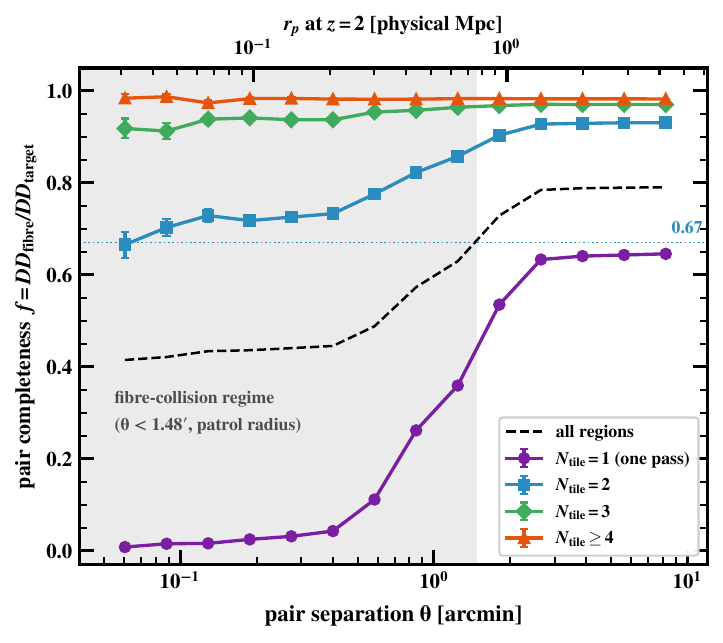}
\caption{Fibre-assignment completeness of quasar pairs, $f=DD_{\rm fibre}/DD_{\rm target}$, as a function of angular separation, for regions covered by one, two, three or at least four tiles. Error bars are binomial. The shaded area marks the fibre patrol radius; the top axis gives $r_p$ at $z=2$. Single-pass regions lose almost all pairs closer than $\sim1\arcmin$.}
\label{fig:completeness}
\end{figure}
\begin{figure}[!ht]
\centering
\includegraphics[width=\columnwidth]{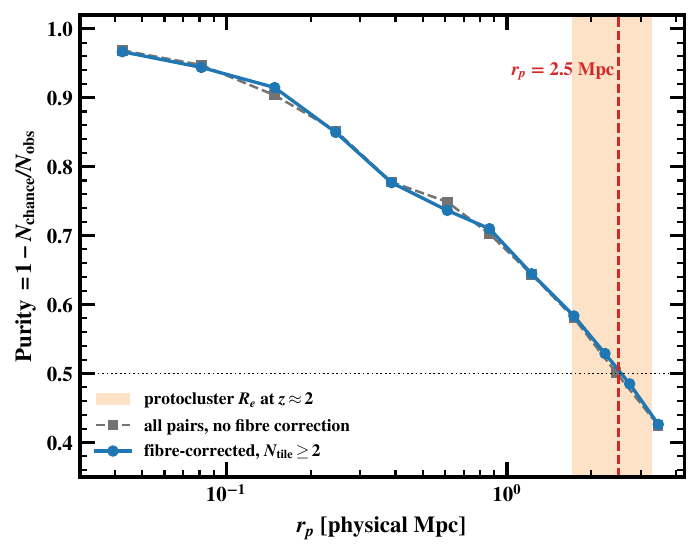}
\caption{Purity of the $|\Delta V|\leq900~\kms$ pair sample as a function of projected separation, from ten redshift-shuffled catalogues. Grey squares: all pairs without fibre correction; blue circles: fibre-corrected pairs with $N_{\rm tile}\geq2$. The dotted line marks 50\%, the dashed line the adopted limit $r_p=2.5$~Mpc, and the shaded band the effective radius of protoclusters at $z\approx2$ \citep{Chiang2013}.}
\label{fig:purity}
\end{figure}
\begin{table*}[!ht]
\caption{Clustering of pair members with all DESI quasars, and of the quasars themselves.}
\label{tab:amp}
\centering
\setlength{\tabcolsep}{4pt}
\begin{tabular}{lrrcccc}
\toprule
Sample & $N_{\rm pairs}$ & $N_{\rm tracers}$ & $A(w_p)$ & $A(\xi_0)$ & $b$ & $\log M_{\rm h}$ \\
\midrule
All quasars (auto) & --- & --- & 1 & 1 & $2.68\pm0.02$ & 12.42 \\
$r_p\leq1$~Mpc & 5352 & 10\,519 & $1.69\pm0.14$ & $1.59\pm0.09$ & $4.53\pm0.36$ & $13.14\pm0.10$ \\
$1<r_p\leq2.5$~Mpc & 13\,915 & 26\,595 & $1.78\pm0.09$ & $1.83\pm0.05$ & $4.77\pm0.24$ & $13.21\pm0.06$ \\
$2.5<r_p\leq4$~Mpc & 19\,169 & 36\,096 & $2.01\pm0.06$ & $2.05\pm0.04$ & $5.39\pm0.17$ & $13.35\pm0.04$ \\
Chance, $r_p\leq2.5$~Mpc & 46\,292 & 82\,383 & $1.08\pm0.05$ & $1.09\pm0.03$ & $2.89\pm0.13$ & $12.54\pm0.07$ \\
\bottomrule
\end{tabular}
\tablefoot{Pairs from the LSS sample with both members in $N_{\rm tile}\geq2$ regions. Physical pairs have $|\Delta V|\leq900~\kms$; chance pairs have $6000<|\Delta V|\leq12\,000~\kms$. $N_{\rm tracers}$ counts each quasar once. Amplitude ratios are measured at 15--40~$\hmpc$ with jackknife errors. The bias is $b=A(w_p)\,b_{\rm q}$, and $M_{\rm h}$ (in $h^{-1}M_\odot$) is the halo mass with that bias in the \citet{Tinker2010} relation at $z_{\rm eff}=1.87$.}
\end{table*}
\paragraph{Comparison with \citet{Jing2026}.}\footnote{\url{https://github.com/astroliang/desi_dr1_qp}} Of the 835 pairs that they classify as physical (QP) at $1.2\leq\bar z\leq2.8$, 588 are in our sample. Of the rest, 79 have a member with \texttt{ZWARN}~$\neq0$ and 168 have $900<|\Delta V|<2000~\kms$ with our redshifts. Their inspection flags 29 of our pairs as lensed-quasar candidates and 121 as contaminants; 63 more are candidates without a confirmed classification, which we keep. Removing the 150 flagged pairs changes the clustering amplitude of the $r_p\leq1$~Mpc bin by less than 0.01.

\begin{table}[!ht]
\caption{Power-law fits to $w_p$ at 10--40~$\hmpc$.}
\label{tab:pl}
\centering
\begin{tabular}{lccc}
\toprule
Sample & $r_0$ [$\hmpc$] & $\gamma$ & $\chi^2/\nu$ \\
\midrule
All quasars & $7.82\pm0.21$ & $2.07\pm0.05$ & 1.9/2 \\
$r_p\leq1$~Mpc & $8.6\pm1.3$ & $1.83\pm0.23$ & 0.9/2 \\
$1<r_p\leq2.5$~Mpc & $10.6\pm0.9$ & $2.13\pm0.22$ & 7.2/2 \\
$2.5<r_p\leq4$~Mpc & $10.78\pm0.36$ & $2.04\pm0.09$ & 0.1/2 \\
Chance & $7.60\pm0.41$ & $1.99\pm0.09$ & 0.8/2 \\
\bottomrule
\end{tabular}
\tablefoot{Errors are scaled by $\sqrt{\chi^2/\nu}$ where $\chi^2/\nu>1$.}
\end{table}

\section{Comparison with SDSS}
\label{app:sdss}
Since MR25 used SDSS, it is worth asking whether DESI gains pairs by covering a larger volume or by sampling the same volume more densely. To compare with MR25 we apply our selection to the SDSS DR16Q catalogue \citep{Lyke2020} with their limit $r_p\leq2$~Mpc; with the MR25 velocity cut instead we recover 3094 SDSS pairs, close to their 2777. DESI finds 28\,656 pairs against 7251, four times more.

Volume plays a minor role. Both samples cover the same redshift range, and only 21\% of the DESI pairs lie outside the SDSS footprint, so the larger area contributes a factor 1.06 (Fig.~\ref{fig:sdss}c). Inside the common $\approx9600$~deg$^2$, DESI has 22\,754 pairs against 6080, 3.7 times more. This follows from the quasar density, 88 against 43~deg$^{-2}$ at $1.2\leq z\leq2.8$, or a mean comoving separation of 41 against 52~$\hmpc$: pair counts scale with the square of the density. The gain is similar at all redshifts (3.3--4.1 in bins of $\Delta z=0.4$) and is 3.3--3.8 at $\theta>62\arcsec$; it rises to 4.5 below the 62\arcsec\ SDSS fibre-collision scale ($r_p\lesssim0.5$~Mpc; Fig.~\ref{fig:sdss}b), thanks to the repeated DESI passes.

The extra quasars are fainter. In the common area, only 11\% of the DESI pairs have both members in DR16Q and 59\% have neither. The members missing from DR16Q have a median $r=21.7$, against 20.6 for those in it, that is, about 2.8 times less luminous at the same redshift (Fig.~\ref{fig:sdss}a). DESI therefore extends the pair sample down the quasar luminosity function rather than to a larger volume. This is the more useful gain for protoclusters: the same structures are marked by four times more pairs, in a census less dominated by the brightest quasars. Faint quasars are not more often in pairs: the fraction of DESI quasars in a pair is $\approx5\%$ at all magnitudes.

\section{Visual and spectroscopic inspection}
\label{app:inspection}
To check that the algorithmic selection returns real quasar pairs, we inspected a random subsample of 30 pairs in each $r_p$ bin. For each member we retrieved the Legacy Survey DR10 $grz$ image\footnote{\url{https://www.legacysurvey.org/viewer}} and the DESI DR1 spectrum through SPARCL,\footnote{\url{https://astrosparcl.datalab.noirlab.edu/}} and automatically tested the spectral class and redshift, the Legacy Survey morphology, the presence of a compact source at the fibre position and of nearby saturated stars. Of the 90 pairs, 80 pass every test and nine more are genuine pairs with a warning: in eight, one member is fainter than $r=22.5$ or classified as slightly extended, and one member lies on the halo of a bright foreground galaxy, which we confirmed by eye. One pair, in the 2.5--4~Mpc bin, contains a galaxy whose redshift was assigned by the quasar afterburners; the contamination of the sample is therefore $\approx1\%$.

\begin{figure}[!t]
\centering
\includegraphics[width=0.90\columnwidth]{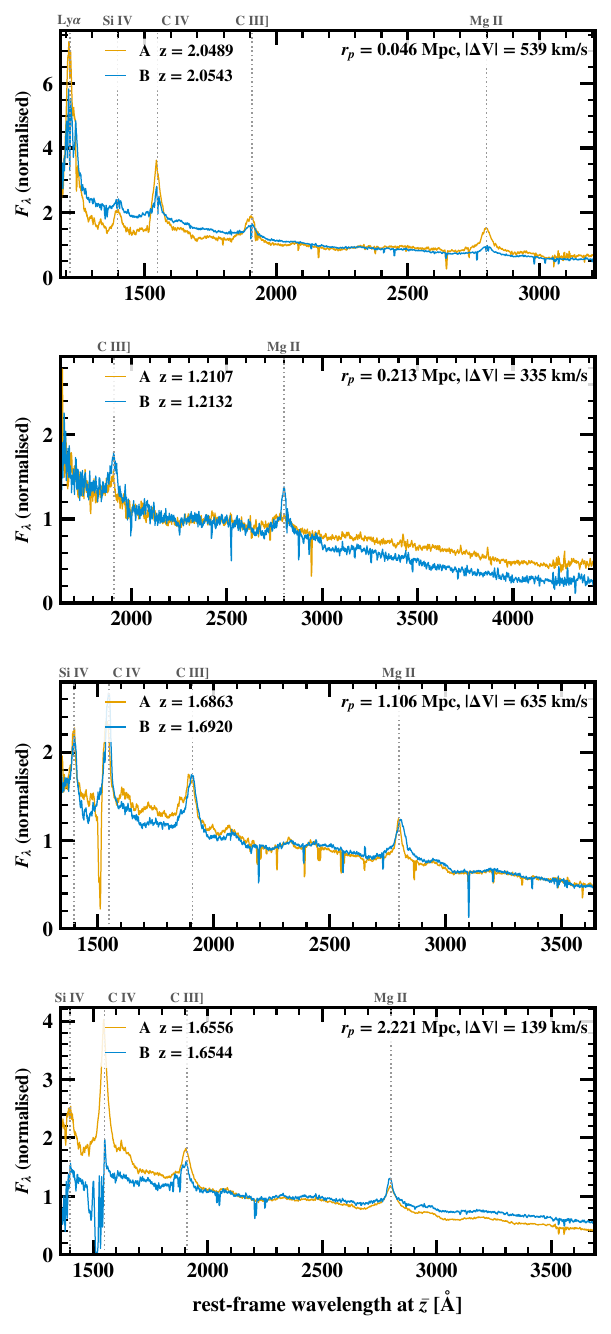}
\caption{DESI DR1 spectra of the two members of each pair in Fig.~\ref{fig:examples}, from top to bottom in order of increasing separation, smoothed over 15 pixels and normalised at rest-frame 1300--3000~\AA. Dotted lines mark the main emission lines at the mean redshift.}
\label{fig:spectra}
\end{figure}

Lensed images of one quasar have identical redshifts and a nearly wavelength-independent flux ratio. Figure~\ref{fig:spectra} shows the spectra of the four pairs of Fig.~\ref{fig:examples}. In all of them the emission lines differ in strength and profile, and the flux ratio between the two members changes by factors of 1.2--2.3 between rest-frame 1450 and 3000~\AA; in the 2.2~Mpc pair one member is a broad absorption line quasar. None of the 90 inspected pairs is closer than 15\arcsec\ with $|\Delta V|<200~\kms$, the regime where a lens could mimic a pair, and the lensed candidates identified by \citet{Jing2026} are already removed (Sect.~\ref{sec:pairs}).

Finally, we inspected 30 quasars without a companion in $N_{\rm tile}\geq2$ regions that have a quasar target without a fibre within 0.5~Mpc. In 27 a source is detected at the target position, 23 of them as compact as the quasars: plausible missed companions, rare in $N_{\rm tile}\geq2$ regions.

\end{appendix}

\end{document}